\documentstyle[twocolumn,prl,aps,epsfig,floats]{revtex}
\begin{document}
\draft
%%%%%%%%%%%%%%%%%%%%%%%%%%%%%%%%%%%%%%%%%%%%%%
%Title
%Abstract
%%%%%%%%%%%%%%%%%%%%%%%%%%%%%%%%%%%%%%%%%%%%%%%%%%%%%%%%%%%%%%%%%%

%%2col
\twocolumn[\hsize\textwidth\columnwidth\hsize\csname @twocolumnfalse\endcsname
%% start of wide text
%%2col

\newcommand{\ibf}{\mbox{\boldmath $f$}}
\title{
Toward a systematic $1/d$ expansion: Two particle properties
}

\author{
Gergely Zar{\'a}nd$^{1,2}$,  Daniel L. Cox$^1$, and Avraham Schiller$^3$
}

\address{
$^1$Department of Physics, University of California Davis, CA 95616\\ 
$^2$Research Group of the Hungarian Academy of Sciences, Institute of Physics,
TU Budapest, H-1521
$^3$Racah Institute of Physics, The Hebrew University, Jerusalem 91904, Israel
}
\date{\today}
\maketitle

\begin{abstract}
We present a procedure to calculate $1/d$ corrections to the
two-particle properties around the infinite dimensional dynamical
mean field limit. Our method is based on a modified version
of the scheme of  Ref.~\protect{\onlinecite{SchillerIngersent}}.
To test our method    we study the Hubbard model at half filling 
within the  fluctuation exchange approximation  (FLEX), 
a  selfconsistent generalization  of iterative perturbation theory. 
Apart from the inherent unstabilities of FLEX, our method is  stable 
and results in causal solutions.  We find that $1/d$ corrections 
to the local approximation are  relatively  small 
in the Hubbard model.
\end{abstract}

%% 71.10.Hf -- Non-Fermi-liquid ground states, electron phase diagrams
%%             and phase transitions in model systems
%% 71.27.+a -- Strongly correlated electron systems; heavy fermions.
%% 72.15.Qm -- Scattering mechanisms and Kondo effect
%% 75.20.Hr -- Local moment in compounds and alloys; Kondo effect,
%%             valence fluctuations, heavy fermions.

%%2col
%% end of wide text
]
\narrowtext
%%2col
%Section 1
%Introduction
%%%%%%%%%%%%%%%%%%%%%%%%%%%%%%%%%%%%%%%%%%%%%%%%%%%%%%%%%%
During the past few years  dynamical mean field theory (DMFT) 
became one of the most popular methods to study strongly 
correlated systems \cite{Kotliar}. 
DMFT developed from the path-breaking 
observation \cite{MetznerVollhard} that in the limit $d\to\infty$
of a $d$-dimensional lattice model with suitably rescaled hopping 
parameters, spatial fluctuations are completely suppressed
and the self-energy becomes local. As a consequence, the self-energy 
can be written as a functional of the on-site Green's function 
of the electrons and the lattice problem reduces to a quantum impurity
problem, where the impurity is embedded in a selfconsistently
determined environment.
The main virtue of this method is that it 
captures all local {\em time-dependent} correlations 
and makes possible  to study,  e.g. the Mott-Hubbard 
transition or the phase diagram of different Kondo lattices in  
detail.

While in the case of the Mott-Hubbard transition the transition 
seems to be driven by the above-mentioned local fluctuations, in 
many cases correlated  hopping \cite{Avi} or inter-site interaction 
effects \cite{Vlad,intersite} may play a crucial role as well, 
and while some of  these effects can be qualitatively  captured by a  
natural extension  of the DMFT,  others are beyond the scope of it and 
would only appear as $1/d$  corrections. Furthermore,  in order 
to check the quality   of the local approximation for a finite 
dimensional  system of interest, it is very important to compare it with 
the size of the  appearing $1/d$ corrections as well.

Several attempts have been made to partially restore some of 
the spatial  correlations lost in the DMFT. One of the most successful ones 
is the cluster approximation proposed by  Jarrell et al \cite{Jarrell}. 
This method has the advantage of being causal, however, it requires
considerable numerical prowess and it is not systematic in the 
small parameter $1/d$. Another  method based on the systematic expansion 
of the generating functional has been suggested by
Schiller and Ingersent \cite{SchillerIngersent}. However, despite of its technical and conceptual 
simplicity,  this method has not been used very extensively because 
it seemed to be somewhat  unstable and in some cases  gave 
artificial  non-causal  solutions. 

In the present work we first show, that the method of Schiller and 
Ingersent (SI) can be considerably stabilized by a minor, however
crucial change in the algorithm, assuring that the contributions of some 
unwanted spurious diagrams exactly cancel. The price for this 
stability is a somewhat increased computation time, since in 
each cycle of the original algorithm an additional subcycle
is needed to assure cancellation. With this change the SI  
method can then be safely extended  to the calculation 
of two-particle properties. Here the main difficulties  
%%We circumvent the appearing new difficulties 
are connected to the inversion involved in the solution of the 
Bethe-Salpeter equation and the non-locality of the irreducible  
vertex functions.  These difficulties are  cicrcumvented  by  introducing 
bond variables for  the two-particle propagators. Finally, we test the 
general  formalism with  the fluctuation exchange  approximation 
(FLEX) \cite{Bickers}.

Although the method presented applies to arbitrary lattice structures
and various models with nearest neighbor interactions,
for concreteness, let us consider the Hubbard model on a $d$-dimensional 
hypercubic  lattice at half filling:
\begin{equation}
H = {t\over \sqrt{d}} \sum_{<i,j>, \sigma} 
c^{\dagger}_{i\sigma} c_{j\sigma} + U \sum_i (n_{i\uparrow}-{1\over
2}) ( n_{i\downarrow}-{1\over 2})\;.
\end{equation}
Here the dynamics of the conduction electrons $c_{i\sigma}$ is driven 
by the hopping $t$ between nearest neighbor sites,
$n_{i\sigma}  = c^{\dagger}_{i\sigma} c_{i\sigma}$ is the occupation 
number, and the electrons interact via the on-site Coulomb repulsion $U$.

In the SI formalism one considers the following  
single ($n=1$)  and a two-impurity  $(n=2)$ imaginary 
time effective functionals  to generate $1/d$ corrections:  
\begin{eqnarray}
S^{(n)} &=&  \sum_\sigma \sum_{\alpha,\beta = 1}^n 
\int d\tau \int d\tau' {\bar c}_{\alpha \sigma}(\tau) 
\bigl[{\cal G}^{(n)}\bigr]^{-1}_{\alpha,\beta} (\tau-\tau') c_{\beta \sigma}(\tau') \nonumber \\
& +& \sum_{\alpha = 1}^n U \int d\tau n_{\alpha \uparrow}(\tau) 
n_{\alpha \downarrow}(\tau)\;.
\end{eqnarray}
Here, as usually,  ${\bar c}_{\alpha \sigma}(\tau)$ and 
${ c}_{\alpha \sigma}(\tau)$ denote Grassman fields, and the
indices $\alpha$ and $\beta$ label the sites for $n=2$ while
they are redundant for $n=1$.
The 'medium  propagators' ${\cal G}^{(1)}$ and ${\cal G}^{(2)}$ must be
 chosen in such a way that the dressed impurity propagators $G^{(1)}$ and 
$G^{(2)}$ coincide with the full on-site and nearest neighbor 
lattice propagators, $G^{\rm latt}_{00}$ and $G^{\rm latt}_{01}$:
\begin{equation}
G^{(1)} = G^{(2)}_{11} =  G^{\rm latt}_{00}\; \phantom{n},
 \phantom{nnn} G^{(2)}_{12} = G^{\rm latt}_{01}\;.
\end{equation}
In this case one can easily show that --- restricting oneself to   
skeleton diagrams of the order of ${\cal O}(1/d)$ ---
the impurity  self energies $\Sigma^{(1)}$ and $\Sigma^{(2)}_{\alpha\beta}$ 
and the diagonal and off-diagonal lattice self energies, 
$\Sigma^{\rm latt}_0$ and $\Sigma^{\rm latt}_1$  are related by 
\cite{SchillerIngersent}
\begin{eqnarray}
\Sigma^{\rm latt}_0 &=& \Sigma^{(1)} + 2d(\Sigma^{(2)}_{11} - \Sigma^{(1)})\;,
\label{eq:Sigma0} \\
\Sigma^{\rm latt}_1 &=& \Sigma^{(2)}_{12}\;.\label{eq:Sigma12}
\end{eqnarray}
Knowing the lattice self energy the lattice Green function 
can then be expressed  as  
\begin{equation} 
G_{lm}^{\rm latt}(i\omega) = {1\over1 +\sqrt{d} \; \Sigma^{\rm latt}_1}
G^{0}_{lm}\left({i\omega - \Sigma^{\rm latt}_0(i\omega) \over
1 +\sqrt{d} \; \Sigma^{\rm latt}_1(i\omega)}\right)\;,
\end{equation}
where $G^{0}_{lm}(z)$ and $G_{lm}^{\rm latt}(z)$ denote the unperturbed 
and dressed lattice propagators between sites $l$ and $m$, 
respectively.

Based on the relations above SI suggested the following simple iterative 
procedure to obtain a solution that includes  ${\cal O}(1/d)$ corrections: 
$$
{\cal G}^{(1,2)}  \Rightarrow {\Sigma}^{(1,2)} \Rightarrow 
\Sigma^{\rm latt} ,  G^{\rm latt} \Rightarrow  {\cal G}^{(1,2)}\;.
$$ 
A careful analysis shows, however, that the second step in 
this scheme is extremely  unstable. To understand this it is 
enough to notice that the second term 
of Eq.~(\ref{eq:Sigma0}) is constructed in such a way that {\em at the 
fixed point} all  {\em completely local } skeleton diagrams in the 
expansion  of  $\Sigma^{(2)}_{11}$ are canceled by the subtracted 
$\Sigma^{(1)}$ term. 
However, this cancellation only happens under the condition that 
the dressed Green's functions $G^{(2)}_{11}$ and $G^{(1)}$
are exactly the same.
If $G^{(2)}_{11}$ and $G^{(1)}$ differ by a term of ${\cal O}(1/d)$ 
in a given iteration step, the cancellation above
is not exact, and an error  of the order of 
$2d\times {\cal O}(1/d)\sim 1$ is generated immediately. 
Moreover, the generated erroneous term is typically acausal 
because of the subtraction procedure involved in
Eq.~(\ref{eq:Sigma0}), 
and may drive the iteration towards some  more stable
but unphysical fixed point of the integral equations. 
We suggest to replace  the critical steps ${\cal G}^{(1,2)}  
\Rightarrow {\Sigma}^{(1,2)} \Rightarrow  \Sigma^{\rm latt}$ by the 
following procedure: 
(1) Calculate $G^{(2)}$ from ${\cal G}^{(2)}$, 
(2) Determine ${\cal G}^{(1)}$ selfonsistently in such a way that 
${G}^{(1)} \equiv {G}^{(2)}_{11}$ be satisfied, (3) Determine 
$\Sigma^{(1,2)}$ and from them $\Sigma^{\rm latt}$. 
Step (2) above is  crucial  to guarantee that  unwanted terms in 
Eq.~(\ref{eq:Sigma0}) exactly cancel.

%%%%%%%%%%%%%%%%%%%%%%%%%%%%%%%%%%%%%%%%%%%%%%%%%%%%%%%%%%%%%%%%
%Fig. 3
%%%%%%%%%%%%%%%%%%%%%%%%%%%%%%%%%%%%%%%%%%%%%%%%%%%%%%%%%%%%%%%%
\begin{figure}
\begin{center}
\psfig{figure=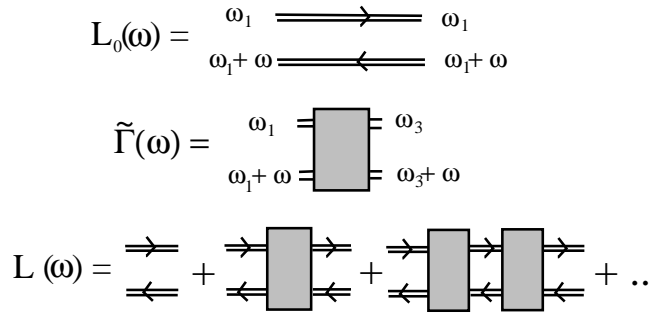,width=8.5cm}
\end{center}
\vspace{0.05cm}
\caption{
\label{fig:bethe}
Graphical representation of the Bethe-Salpeter equation: 
Double lines indicate dressed Fermion propagators, while  
boxes denote particle-hole irreducible vertex diagrams.
}
\end{figure}
%%%%%%%%%%%%%%%%%%%%%%%%%%%%%%%%%%%%%%%%%%%%%%%%%%%%%%%%%%%%%%%%

The two-particle properties can be investigated  in a way similar 
to Ref.~\onlinecite{SchillerIngersent}. To this end we introduce  
the lattice particle-hole irreducible vertex  function
${\bf \tilde \Gamma}^{\rm latt}$, which is connected to the 
full lattice propagator ${\bf L}^{\rm latt} $ by the Bethe-Salpeter 
equation (see Fig.~\ref{fig:bethe}): 
\begin{equation}
{\bf L}^{\rm latt}(\omega) = {\bf L}_0^{\rm latt}(\omega)  ( {\bf 1} + 
{\bf \tilde \Gamma}^{\rm latt}(\omega)  {\bf L}_0^{\rm latt}(\omega) )^{-1},
\label{eq:bethe}
\end{equation}
where $\omega$ denotes the transverse frequency and 
a tensor notation has been introduced in the spatial, spin and
other frequency indices:
$L(\omega)^{\sigma_1,\sigma_2; \sigma_3,\sigma_4}_{ i_1,i_2,\omega_1; 
i_3,i_4,\omega_3}
\to {\bf L}(\omega)$. The 'vertex-free' propagator 
$ {\bf L}_0^{\rm latt}(\omega)$ is defined as 
$L_0^{\rm latt}(\omega)^{\sigma_1,\sigma_2; \sigma_3,\sigma_4}_{ i_1,i_2,
\omega_1; i_3,i_4,\omega_3} =$  $\delta_{\sigma_1\sigma_3}
\delta_{\sigma_2\sigma_4} \delta_{\omega_1\omega_3}$
$G_{i_1,i_3}^{\rm latt}(i\omega_1)$ $G_{i_4,i_2}^{\rm latt}(i(\omega_1+\omega))$.

A detailed analysis shows that up to 
$1/d$ order  the only non-zero matrix elements of 
${\bf \tilde \Gamma}^{\rm latt}(\omega)$ are those 
where the indices $ i_1,i_2,i_3,i_4$ belong to the same or 
two nearest neighbor lattice sites, i.e. a bond.  
A thorough investigation  of the corresponding skeleton diagrams 
shows that ${\bf \tilde \Gamma}^{\rm latt}(\omega)$
can be expressed similarly to 
Eqs.~(\ref{eq:Sigma0}) and (\ref{eq:Sigma12}) as:
\begin{equation}
{\bf \tilde \Gamma}^{\rm latt} = 
\left\{
\mbox{
\begin{tabular}{ll }
${\bf \tilde \Gamma}^{(1)} + 2d ({\bf \tilde \Gamma}^{(2)} - {\bf \tilde \Gamma}^{(1)} )$ &
\phantom{n} $i_1=i_2=i_3 = i_4$, \\
${\bf \tilde \Gamma}^{(2)}$ &
\phantom{n} $i_k\in$ bond, \\
0 & \phantom{n}otherwise, 
\end{tabular}
}\right. 
\label{eq:Gamma_latt}
\end{equation}
where in the second line it is implicitely  assumed that 
the $i_k$'s are not all equal. 
Here the particle-hole irreducible one- and two-impurity 
vertex functions,   ${\bf \tilde \Gamma}^{(1)}$, 
and ${\bf \tilde \Gamma}^{(2)}$ are defined similarly to 
${\bf \tilde \Gamma}^{\rm latt}$, and satisfy the impurity
Bethe-Salpeter  equation:
 \begin{equation}
{\bf L}^{(n)}(\omega) = {\bf L}_0^{(n)}(\omega)  ( {\bf 1} + 
{\bf \tilde \Gamma}^{(n)}(\omega)  {\bf L}_0^{(n)}(\omega) )^{-1},
\label{eq:bethe12}
\end{equation}
with $n=1,2$. Of course, in the impurity case the spatial indices
of the propagators ${\bf L}_0^{(n)}(\omega) $ and ${\bf L}^{(n)}(\omega)$ 
are restricted to the impurity sites, but apart from this
the ${\bf L}_0^{(n)}(\omega) $'s are equal to the lattice 
propagator ${\bf L}_0^{\rm latt}(\omega)$. 

From the considerations above it immediately follows that 
the $1/d$ corrections to the two-particle properties can be 
calculated in the following way: (1) Find the solution of the 
single particle iteration scheme, (2) Determine the one- and two-impurity 
correlators, (3) Invert Eq.~(\ref{eq:bethe12}) to obtain  
${\bf \tilde \Gamma}^{(1)}$ and ${\bf \tilde \Gamma}^{(2)}$, 
(4) Calculate ${\bf \tilde \Gamma}^{\rm latt}$ using 
Eq.~(\ref{eq:Gamma_latt}), and 
(5) Solve the Bethe-Salpeter equation (\ref{eq:bethe}) for
 ${\bf L}^{\rm latt}$ and calculate two-particle response functions from it.

The major difficulties in the procedure above are  associated 
with the inversion appearing in  Eqs.~(\ref{eq:bethe}), since
the  propagator ${\bf L}_0^{\rm latt}$ connects any four lattice 
sites and has an infinite number of frequency indices. The first 
difficulty can be resolved by observing that ${\bf \tilde \Gamma}^{\rm latt}$ 
connects only neighboring sites. Therefore with a introduction of 
{\em bond variables} a partial Fourier transformation can be carried out in 
these, and summations over all pairs of lattice sites 
reduce to a summation over $d$  'bond direction indices' 
and two additional indices specifying the position of the 
electron and the hole within a given bond. 
Furthermore, to avoid overcounting, the vertex function
${\bf \tilde \Gamma}^{\rm latt}$ must be replaced by  a slightly 
modified 'bond vertex function' \cite{ZarCoxScill2}.
A further reduction of the 
matrices  involved can be achieved by diagonalizing the 
propagators in the spin labels. Finally, to carry out the 
summations and inversions over the infinite omega variables
we introduced a frequency cutoff $\omega_c$ and extrapolated 
the $\omega_c=\infty$ result from a finite size scaling analysis 
in this cutoff \cite{Pruschke}, thereby reducing the numerical error of
our calculations below one percent.

%%%%%%%%%%%%%%%%%%%%%%%%%%%%%%%%%%%%%%%%%%%%%%%%%%%%%%%%%%%%%%%%
%Fig. 1
%%%%%%%%%%%%%%%%%%%%%%%%%%%%%%%%%%%%%%%%%%%%%%%%%%%%%%%%%%%%%%%%
\begin{figure}
\begin{center}
\psfig{figure=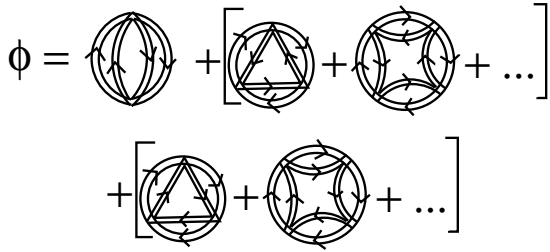,width=7.2cm}
\end{center}
\vspace{0.05cm}
\caption{
\label{fig:diagrams}
The generating functional for FLEX: Double lines denote  
single- and two-impurity dressed cavity propagators $G^{(1)}(\tau,\tau')$
and $G^{(2)}(\tau,\tau')$. Position and spin indices are not 
shown.  The corresponding self-energy diagrams  are obtained as
$\Sigma(\tau-\tau') = -\delta \phi[G(\tau,\tau')] / \delta G(\tau',
\tau)$, and the particle-hole irreducible  vertex is given by 
a similar  second order functional derivative.
}
\end{figure}
%%%%%%%%%%%%%%%%%%%%%%%%%%%%%%%%%%%%%%%%%%%%%%%%%%%%%%%%%%%%%%%%

To test the procedure above one is tempted to try to generalize the
iterative perturbation theory (IPT) applied remarkably successfully
for the $d=\infty$ case \cite{IPT}, however, it is clear from the
discussion above that within  IPT it is impossible to satisfy the
condition ${G}^{(1)} \equiv {G}^{(2)}_{11}$ (which explains why
earlier attempts to generalize  IPT to order $1/d$  failed
\cite{Georges}). We therefore applied the so-called
fluctuation exchange approximation (FLEX) \cite{Bickers}. While this
method is unable to capture the metal insulator transition, it is
able to reproduce the Kondo resonance in the metallic
phase\cite{BickersSC}, has been successfully used to calculate
weak and intermediate coupling properties of the 2-dimensional
Hubbard model \cite{BickersSC}, and it has the important property of
being formulated in terms of the {\em dressed} single particle
Green's functions.  In this approach the interactions between
particles are mediated by fluctuations in the particle-particle and
particle-hole channels, and the self-energies and the particle-hole
(particle-particle) irreducible vertex functions are generated from
the generating $\Phi$ functionals built in terms of the dressed
Green's functions, depicted in Fig.~\ref{fig:diagrams}. A further
advantage of FLEX is that due to the special structure of the
diagrams involved a fast Fourier transform algorithm can be exploited
to increase the speed and  precision of the calculation
substantially.

%%%%%%%%%%%%%%%%%%%%%%%%%%%%%%%%%%%%%%%%%%%%%%%%%%%%%%%%%%%%%%%%
%Fig. 2
%%%%%%%%%%%%%%%%%%%%%%%%%%%%%%%%%%%%%%%%%%%%%%%%%%%%%%%%%%%%%%%%
\begin{figure}
\begin{center}
\psfig{figure=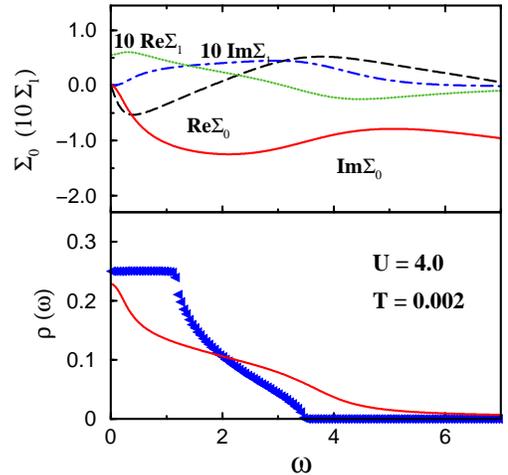,width=6.5cm}
\end{center}
\vspace{0.05cm}
\caption{
\label{fig:self}
Top: Calculated diagonal and off-diagonal lattice self-energies 
for $U = 4$ and $T = 0.002$. All energies are measured in units of 
$t$. Bottom: Local spectral functions for $U=0$ and $U=4$ at 
$T = 0.002$. 
}
\end{figure}
%%%%%%%%%%%%%%%%%%%%%%%%%%%%%%%%%%%%%%%%%%%%%%%%%%%%%%%%%%%%%%%%

\begin{figure}
\begin{center}
\psfig{figure=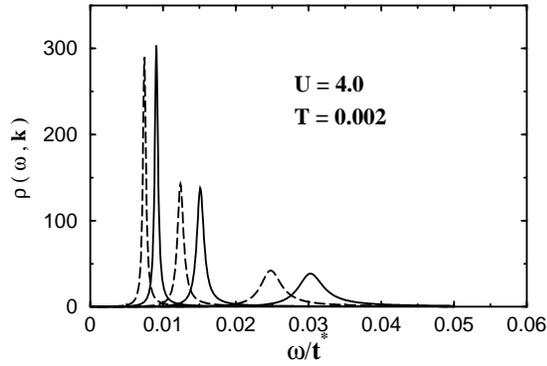,width=7.2cm}
\end{center}
\vspace{0.05cm}
\caption{
\label{fig:rhok}
Momentum dependent spectral functions along the line  
$k = \alpha (\pi,\pi,\pi)$ within the local (dashed lines) and 
$1/d$ calculations (continuous lines). The alpha values 
used were $\alpha = 0.497$, $\alpha = 0.495$, and $\alpha = 0.49$
from left to right. The energy of the quasiparticles has been 
renormalized approximately by a factor of four compared to the 
bare electron energies.
}
\end{figure}

The calculated three-dimensional  diagonal and off-diagonal 
lattice self-energies  are shown in Fig.~\ref{fig:self} together with 
the local spectral  function. These have been obtained 
by means of a Pade approximation to carry out the  
analytic  continuation  from  the  imaginary to the real axis. 
Though in the spectral  function a well-developed Kondo peak 
is observed, 
the FLEX is unable to  reproduce the depletion of spectral weight 
in the neighborhood of it
due to the 'over-regularization'
characteristic to most selfconsistent perturbative schemes.
Remarkably, we experienced no convergence problems similar 
to  those of Ref.~\onlinecite{SchillerIngersent}, apart 
from the ones inherent in FLEX \cite{Bickers}. We checked that the 
spectral  functions integrate to one within numerical precision
and the solutions obtained are  causal. The typical values of 
$\Sigma^{\rm latt}_1$ are nearly an order of magnitude 
smaller \cite{sig_1footn} than  $\Sigma^{\rm latt}_0$, indicating  that 
the local approximation gives surprisingly good results 
and $1/d$ corrections are indeed small as anticipated in 
Ref.~\onlinecite{Kotliar} and also in agreement with  the results of
Ref.~\onlinecite{SchillerIngersent}.  To get some further information about 
the  quality of  local approximation in Fig.~\ref{fig:rhok}
we plotted the momentum dependent spectral functions 
at different points of the Brillouin zone. The $1/d$ 
contributions give typically a 10-20 percent correction, 
but none of the generic properties is modified in the 
paramagnetic phase.

%%%%%%%%%%%%%%%%%%%%%%%%%%%%%%%%%%%%%%%%%%%%%%%%%%%%%%%%%%%%%%%%
%Fig. 4
%%%%%%%%%%%%%%%%%%%%%%%%%%%%%%%%%%%%%%%%%%%%%%%%%%%%%%%%%%%%%%%%
%%%%%%%%%%%%%%%%%%%%%%%%%%%%%%%%%%%%%%%%%%%%%%%%%%%%%%%%%%%%%%%%

\begin{figure}
\begin{center}
\psfig{figure=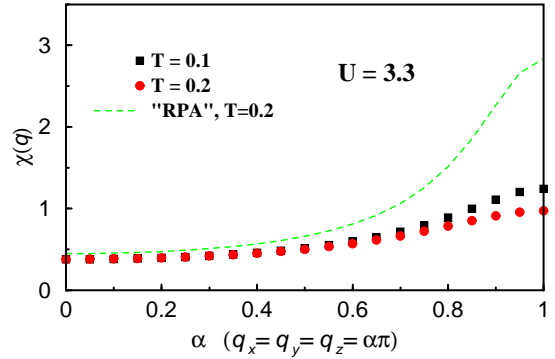,width=7.2cm}
\end{center}
\vspace{0.05cm}
\caption{
\label{fig:chi}
Momentum dependent susceptibilities. The susceptibility 
develops a peak at $(\pi,\pi,\pi)$ as a precursor of the 
antiferromagnetic phase transition.
}
\end{figure}

Once convergence is reached at the single particle level, 
one can turn to the two-particle properties.
Within FLEX this is somewhat simpler, because ---
although many rather complicated diagrams are generated \cite{Bickers,ZarCoxScill2}
---  ${\bf \tilde \Gamma}^{(n)}$  can be built up  directly in terms of
the full lattice Green's functions. We find that similarly to
the off-diagonal self-energy the off-diagonal 
elements of  ${\bf \tilde \Gamma}^{\rm latt}$ are rather small.
Having solved Eq.~(\ref{eq:bethe}) one can calculate various 
correlation functions.
As an example, in Fig.~\ref{fig:chi} we show the momentum dependent 
susceptibility  of the half-filled  Hubbard model in its paramagnetic 
phase for two different temperatures along the $(1,1,1)$
direction, obtained from the FLEX calculations. The susceptibility 
develops  a peak at $(\pi,\pi,\pi)$
at low temperatures, as a sign of  unstability 
toward antiferromagnetic phase transition.

We also determined the  transition 
temperature at several values of $U$ and compared our results with 
existing Monte Carlo data\cite{Scalettar}. We found  
a critical temperature $T_c$  typically by a factor of three lower  
than that of  Ref.~\onlinecite{Scalettar}. This difference is a result 
of the overregularization of the interaction vertex by FLEX.
 Indeed, replacing $ {\bf \tilde \Gamma}^{\rm latt}(\omega)$ by the  
{\em bare} particle-hole  vertex in Eq.~(\ref{eq:bethe}) the order-parameter
fluctuations become larger (see Fig.~\ref{fig:chi}) and $T_c$ is in 
excellent agreement with the Monte Carlo data.

In conclusion, we presented an extended  version of the SI method 
to calculate $1/d$ corrections to the two-particle properties. 
We tested the  new procedure by FLEX. No convergence problems 
and no violation of causality appeared in our method, although 
this is not generally guaranteed within the present scheme. 
Our method  should be  tested on other models
and with other, more time-consuming methods in the future 
as well.

%%%%%%%%%%%%%%%%%%%%%%%%%%%%%%%%%%%%%%%%%%%%%%%%%%%%%%%%%%%%%%%%
%Fig. 5
%%%%%%%%%%%%%%%%%%%%%%%%%%%%%%%%%%%%%%%%%%%%%%%%%%%%%%%%%%%%%%%%
%%%%%%%%%%%%%%%%%%%%%%%%%%%%%%%%%%%%%%%%%%%%%%%%%%%%%%%%%%%%%%%%

The authors are grateful to  N. Bickers for valuable  discussions. 
This research has been supported by  
the U.S - Hungarian Joint Fund No. 587, grant No. DE-FG03-97ER45640 of the
U.S DOE Office of Science, Division of Materials Research, 
and Hungarian Grant Nr. OTKA T026327, OTKA F29236, and 
OTKA T029813.


\begin{thebibliography}{99}
\bibitem{sig_1footn} In reality, $\Sigma_1^{\rm latt}$ appears with a 
prefactor $\sqrt{d}$ in the Green's function which, however, does not
change this conclusion.
\end{thebibliography}
\end{document}